\documentstyle[amsbsy,amstex,aps,12pt,epsf]{revtex}

\def\eqref#1{Eq.~(\ref{#1})}

\def\phi{\varphi}

\def\({\left(}
\def\){\right)}
\def\[{\left[}
\def\]{\right]}
\def\<{\left\langle}
\def\>{\right\rangle}
\def\<{\left\langle}
\def\>{\right\rangle}

\def\bea{\begin{eqnarray}}
\def\eea{\end{eqnarray}}

\makeatletter

\def\vlp{\mathopen{\boldsymbol{(}}}    
\def\vrp{\mathclose{\boldsymbol{)}}}   

\title{On Batchelor passive advection by a finite-time correlated 
random velocity field}
\author{Stanislav~A.~Boldyrev\\
{\em Institute for Theoretical Physics, 
Santa Barbara, California 93106}}

\date{\today}

\begin{document}

\input psfig.sty


\maketitle

\begin{abstract}
\vskip5mm
The Batchelor passive advection is an advection by a smooth velocity
field. If the velocity field is a $\delta$-correlated in time 
random Gaussian process, then the problem is reduced to quantum
mechanics of fluctuating velocity 
gradient~$\frac{\partial u^i}{\partial x^j}(t)$. 
For the finite-time correlated velocity field, such a reduction
does not exist.  To illustrate this point, the second moment of 
a passively advected magnetic field is considered, and the 
stochastic calculus is used to find finite-time corrections 
to its growth rate. The growth rate depends on large scale
properties of the velocity field. Moreover, the problem is not 
universal with respect to the short-time regularization: different 
regularizations give different answers for the growth rate.

\end{abstract}

\section{Introduction}
Magnetic field passively advected by an incompressible 
velocity field obeys the induction equation
\begin{eqnarray} 
\partial_t B^i + v^k\partial_k B^i - B^k \partial_k v^i=\eta \Delta B^i,
\label{induction}
\end{eqnarray}
where~$\eta$ is the diffusivity. We will assume that the
diffusivity is negligibly small (large Prandtl number), 
and the magnetic field is therefore frozen
into the fluid. The incompressibility of the velocity field is not
relevant for further calculations; for the compressible case,
Eq.~(\ref{induction}) holds for~$B^i/\rho$, where~$\rho$ is the
density of the fluid. 
The random velocity field is assumed to be Gaussian in the Eulerian frame:
\begin{eqnarray}
\label{correlator}
&{}&\langle v^i({\bf x},t)\rangle=0, \nonumber \\
&{}&\langle v^i({\bf x},t)v^j({\bf x}',t') \rangle 
=\kappa^{ij}({\bf x}-{\bf x}', t-t'),
\end{eqnarray}
where $\kappa^{ij}({\bf x},t)$ is an arbitrary function 
regular at~${\bf x}=0$, this is the so-called Batchelor
regime~\cite{Batchelor}. 
The solution of the Eq.~(\ref{induction}) is
\begin{eqnarray}
B^i(x,t)= B^i_0 +\int\limits^t_0
\frac{\partial {\tilde v}^i(y,\tau)}{\partial y^k} \,\mbox{d}\tau \, 
B^k_0
= B^i_0 + 
\int\limits^t_0 \frac{\partial {\tilde v}^i \vlp y(x, t),\tau \vrp }{\partial x^k}\, 
\mbox{d}\tau \, B^k\vlp x(y,t),t \vrp ,
\label{differential}
\end{eqnarray}
where $y^i=x^i(y,0)$ and $B^i_0=B^i(y,0) $ are initial values of 
the fluid-particle coordinate and of the magnetic field. The 
object~${\tilde
v}^i (y, \tau) $ in the integrand 
is the {\em Lagrangian} velocity field. 
The {\em Eulerian} velocity has the form~$v^i(x,\tau)={\tilde
v}^i \vlp y(x, \tau ),\tau \vrp $. To make use of the
formula~(\ref{correlator}), we need to ``stop" the moving
point~$x^i=x^i(y,t)$ in the function~${\tilde
v}^i \vlp y(x, \tau ),\tau \vrp$. This is done
differently in the Eulerian and Lagrangian frames~(see
Fig.\ref{trajectories}). 

\section{Lagrangian frame}
\label{lagrangian_frame}
In the Lagrangian frame,~${\bf y}$ is fixed and we need to 
express~${\tilde v}^i (y, \tau )$ in terms of~$v^i(y,\tau)$. The
general expression 
is~${\tilde v}^i ( y, \tau )=
v^i\left( y^k+\int\limits^{\tau}_0{\tilde v}^k (y, \tau' ), \tau \right)$. For
small~$t$, we expand this expression up to the second order in~
$\int\limits^{\tau}_0{\tilde v}^i (y, \tau' )$, and iterate once with
respect to~$\tilde v$. We obtain

\begin{eqnarray}
{\tilde v}^i(\tau)=
 v^i(\tau)+v^i_l(\tau)\int\limits^{\tau}_0 v^l (\tau ')+ 
v^i_l(\tau)\int\limits^{\tau}_0 v^l_j (\tau ')\int\limits^{\tau'}_0 
v^j(\tau'') +
\frac{1}{2}v^i_{lj}(\tau)\int\limits^{\tau}_0 v^l (\tau
')\int\limits^{\tau}_0 
v^j(\tau'') +\dots ,
\label{velocity_expansion_lagrange}
\end{eqnarray}
where all the velocities are taken at the point~${\bf y}$, 
and we use the short-hand notation~$v^i_l(\tau)\equiv \partial
v^i(y, \tau)/\partial y^l$ and~$v^i_{lj}(\tau)\equiv \partial^2
v^i(y, \tau)/\partial y^l\partial y^j$. 

To obtain the equation for the
second moment of~$B^i$, we substitute the
expansion~(\ref{velocity_expansion_lagrange}) into
the formula~(\ref{differential}), raise the latter to 
the second power
and average using formula~(\ref{correlator}).  
We assume that~$t$ is much smaller than
the inverse growth rate of the second moment, and that the correlation
time~$\tau_c$ is much smaller than~$t$. Assuming
isotropic and spatially homogeneous initial distribution, 
$\langle B^i_0 \rangle =0$, $\langle
B^i_0B^j_0 \rangle=\frac{1}{d}\delta^{ij} H_2(0)$, we find the 
equation for the second-order 
moment~$H_2(t)=\langle |{\bf B}|^2\rangle (t)$ in the Lagrangian 
frame. For the incompressible velocity field, it takes the form

\begin{eqnarray}
\label{growth_lagrangian}
\frac{\mbox{d}}{\mbox{d}t} H_2=-\frac{1}{d}\kappa^{ii}_{jj}H_2
-\frac{1}{24d}\tau_c(\kappa^{ii}_{jjlm}\kappa^{lm}-8\kappa^{im}_{jl}
\kappa^{il}_{jm}
+10\kappa^{ii}_{lm}\kappa^{lm}_{jj})H_2.
\end{eqnarray}

\noindent To derive this equation we assumed that
$\kappa(y,t)\simeq\kappa(y)T(t)$, where
$T(t)$~is a $\delta$~function smeared over the correlation 
time~$\tau_c$. In particular, we have chosen the ``box"
regularization:~$T(t)=1/\tau_c$ for~$t \in [-\tau_c/2, \tau_c/2]$,
and~$T(t)\equiv 0$ otherwise. The coefficients in the correction
terms in the formula~(\ref{growth_lagrangian}) are given by the 
following integrals:

\begin{eqnarray}
\label{I_1}
I_1&=&\frac{1}{2}\int\limits^t_0 \int\limits^t_0 
T(\tau_1-\tau_2)
\int\limits^{\tau_1}_{\tau_2}
\int\limits^{\tau_2}_{\tau_1}
T(\tau'-\tau'')=-\frac{1}{24}\,t\tau_c +\dots, \\
\label{I_2}
I_2&=& 2\int\limits^t_0 \int\limits^t_0 \int\limits^{\tau_1}_0
\int\limits^{\tau_2}_0
T(\tau_2-\tau')T(\tau_1-\tau'')=\frac{1}{3}\,t\tau_c +\dots, \\
\label{I_3}
I_3&=&\int\limits^t_0 \int\limits^t_0 T(\tau_1-\tau_2)
 \int\limits^{\tau_1}_0
\int\limits^{\tau_2}_0
T(\tau'-\tau'')=\frac{t^2}{2}-\frac{5}{12}\,t\tau_c +\dots. 
\end{eqnarray}

\noindent All the derivatives of~$\kappa(y)$ in 
formula~(\ref{growth_lagrangian}) are taken at~$y=0$; the subscripts
denote derivatives with respect to the corresponding components
of~$\bf y$, and~$d$ is the space dimension.

One can easily check that the first-order $\tau_c$~correction to 
the growth rate is negative. It is important to note that 
this correction is not universal since the 
integrals~(\ref{I_1}-\ref{I_3}) depend on
the chosen regularization~$T(t)$ of the  $\delta$~function.

\newpage
\section{Eulerian frame}
\label{eulerian_frame}

In the Eulerian frame, the point~${\bf x}$ is fixed in
the formula~(\ref{differential}), and we need
to express~${\tilde
v}^i \vlp y(x, t ),\tau \vrp$ in terms of~$v^i(x,\tau)$.
By analogy with the expression~(\ref{velocity_expansion_lagrange}), we
write~${\tilde v}^i(y, \tau)=v^i 
\left(x^k-\int\limits^{t}_{\tau}{\tilde v}^k (y, \tau' ), \tau \right)$.  The
expansion now takes the form
\begin{eqnarray}
{\tilde v}^i(\tau)=
 v^i(\tau)-v^i_l(\tau)\int\limits^{t}_{\tau} v^l (\tau ')+ 
v^i_l(\tau)\int\limits^{t}_{\tau} v^l_j (\tau
')\int\limits^{t}_{\tau'} 
v^j(\tau'') +
\frac{1}{2}v^i_{lj}(\tau)\int\limits^{t}_{\tau} v^l (\tau
')\int\limits^{t}_{\tau} 
v^j(\tau'')+\dots,
\label{velocity_expansion_euler}
\end{eqnarray}
where we use the short-hand notation~$v^i_l(\tau)\equiv \partial
v^i(x, \tau)/\partial x^l$.
To calculate the second moment of~$B^i$, we have to use the 
expression~(\ref{differential}):

\begin{eqnarray}
\label{derivative_eulerian}
B^i(x,t) =B^k_0 \left( \delta^i_k -\int \limits^t_0 
\frac{\partial{\tilde v}^i(y,\tau)}{\partial x^k}
\,\mbox{d}\tau\right)^{-1},
\end{eqnarray}
that should be expanded up to the fourth order in~$\int \limits^t_0 
\frac{\partial{\tilde v}^i(y,\tau)}{\partial x^k}
\,\mbox{d}\tau $. Then, we substitute the 
expansion~(\ref{velocity_expansion_euler}) for~${\tilde v}$. 
Finally, assuming isotropic and spatially homogeneous initial distribution, 
$\langle B^i_0 \rangle =0$, $\langle
B^i_0B^j_0 \rangle=\frac{1}{d}\delta^{ij} H_2(0)$, and 
making use of~(\ref{correlator}), we can
find the equation for~$H_2(t)$. We do not present the detailed
calculation here since for the incompressible velocity field,
the answer coincides with the Lagrangian
case~(\ref{growth_lagrangian}), as it should.

The appearance of the 
term~$\kappa^{ii}_{jjlm}\kappa^{lm}$ in Eq.~(\ref{growth_lagrangian}) 
shows that for the finite correlation
time~$\tau_c$, the velocity field~$v^i(x,t)$ cannot be
treated as linear in the calculation of the growth rate. This is
obvious since a Lagrangian particle is swept by the finite
distance~$v^i\tau_c$ before the velocity field becomes decorrelated. 
The appearance of~$\kappa(0)$ (or, equivalently,~$u_{rms}$) is a 
manifestation of the absence of the Galilean invariance. For 
the velocity field with finite correlation time, such invariance is
broken as it can be seen from Eq.~(\ref{correlator}).

{

\begin{figure} [tbp]
\centerline{\psfig{file=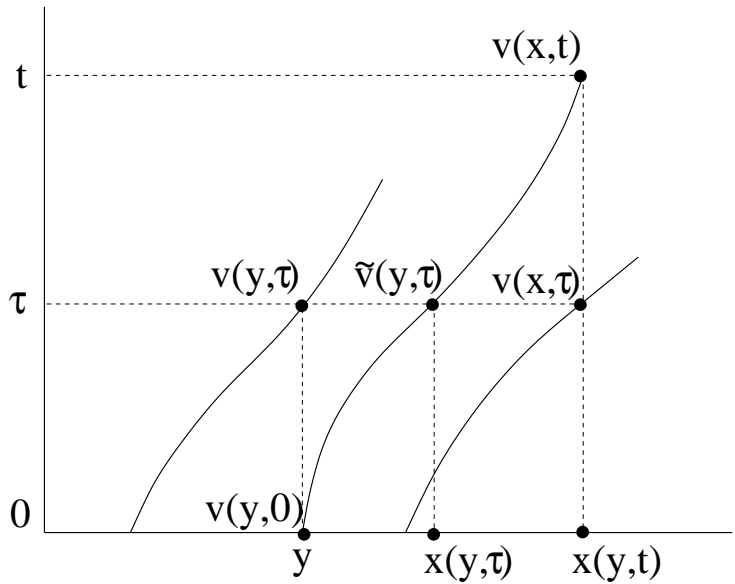,width=3.8in}}
\caption{Trajectories of Lagrangian particles.}
\label{trajectories}
\end{figure}
}

\section{Discussion}
\label{discussion}
The Batchelor advection, i.e. the advection by a smooth
velocity field is a good model for many physical problems. 
It is usually a valid approximation when one is interested in
the correlators of the advected fields on the scales $l_{adv}$ 
that are much smaller than the viscous scale of the velocity
field~$l_v$. 

An additional simplification arises when the correlation time of the
velocity field~$\tau_c$ is much smaller than the inverse growth 
rates of the advected
fields. Mathematically, such a limit is described by the
$\delta$-correlated in time velocity field. In such a case, the growth
rates depend only on~$\kappa''(0)$, as a consequence of the scale
invariance. In this case one can substitute 
the linear velocity field~$v^i(x,t)=\sigma^i_k(t)x^k$, $\langle
\sigma^i_l(t) \sigma^j_m(0) \rangle=\kappa^{ij}_{lm}(0)\delta (t)$,
for the real field, and thus reduce the field-theoretical
problem to the quantum mechanics of the fluctuating 
matrix~$\sigma^i_k$(t); for details and references, 
see~\cite{Son,Balkovsky_Fouxon,Chertkov_etal_dynamo}. 

We have demonstrated that the possibility of such
a reduction is an artifact of the $\delta$-correlated velocity field,
and have presented a simple method for calculating the finite-time
corrections to the growth rates. The criterion of the 
applicability of the quantum mechanical reduction is not only
$l_{adv}/l_v \ll 1$, but also~$\tau_c \, \partial v/{\partial x}\ll
1$. For~$\tau_c\neq 0$, the universal growth rates do not exist, they
are determined by the statistics of the velocity field on the 
integral scale and by the form of the 
short-time regularization. This fact is 
sometimes overlooked in the literature; in the 
present note we have tried to clarify the question. 
The general formalism allowing to systematically find the
finite-time corrections to the dynamo growth rates, has been 
developed in~\cite{Schekochihin_Kulsrud} by a different method. A
simple discussion of the stochastic calculus applied to 
the dynamo problem can be found in~\cite{Zeldovich}.
\vskip5mm

I am grateful to Alexander~Schekochihin for many valuable discussions
and comments on both the substance and the style of the paper,  
and to Michael~Chertkov for important conversations.




\end {document}